\documentclass{article}
\usepackage[utf8]{inputenc}
\usepackage{authblk}
\usepackage{indentfirst}
\usepackage{hyperref}
\usepackage{longtable}
\usepackage{setspace}
\usepackage[margin=1in]{geometry}
\usepackage{graphicx}
\graphicspath{ {./figures/} }
\usepackage{subcaption}
\usepackage{amsmath}
\usepackage{lineno}
\usepackage{multirow}
\usepackage{color,soul}
\usepackage[version=3]{mhchem}

\usepackage[biblabel]{cite}

\title{Phase stability and ionic transport in post-spinel CaV$_2$O$_4$ cathode}

\author[1]{Dereje Bekele Tekliye}
\author[1]{Javeed Ahmad Dari}
\author[1,*]{Gopalakrishnan Sai Gautam}

\affil[1]{Department of Materials Engineering, Indian Institute of Science, Bengaluru, 560012, India}
\affil[*]{Email: \href{mailto:saigautamg@iisc.ac.in}{saigautamg@iisc.ac.in}}

\date{}

\begin{document}

\maketitle

%%%%%% Abstract %%%%%%
\begin{abstract}
\noindent Calcium-ion batteries (CBs) represent a sustainable and high-energy-density alternative to state-of-the-art lithium-ion technology, yet their advancement is limited by the lack of high-performance intercalation cathodes. Identified via computational screening, post-spinel CaV$_2$O$_4$ has emerged as a promising candidate, though its practical application is hindered by limited electrochemical capacity. To rationalize the limitations associated with CaV$_2$O$_4$, we investigate the thermodynamic and ionic transport characteristics of Ca$_x$V$_2$O$_4$ ($0 \leq x \leq 1$) in this work. By integrating the cluster expansion formalism with Monte Carlo simulations and density functional theory based calculations, we construct the temperature-composition phase diagram of Ca$_x$V$_2$O$_4$, which reveals a complex phase behavior with Ca (de)intercalation consisting of several single phases ($\alpha$ through $\zeta$) that can appear under electrochemical operating conditions at different voltages versus Ca/Ca$^{2+}$. Importantly, we observe the formation of the $\varepsilon$ phase at $x \sim 0.83$ across a temperature window of 370-590~K via invariant reactions, which agrees with observations of subtle slope changes in the experimental voltage profiles. Further, migration barrier calculations confirm that Ca mobility is severely impeded within the $\alpha$ ($x \sim 0$) and $\gamma$ ($x \sim 0.5$) phases, with the strong Ca-vacancy ordering contributing to the high barrier in the $\gamma$ phase. Given the persistent two-phase region stretching across the $\delta$ ($x \sim 0.67$) and the $\gamma$ phases, and the poor Ca mobility in the $\gamma$ phase, we expect the accessible electrochemical capacity in the CaV$_2$O$_4$ system to be kinetically limited to at most half the theoretical capacity at 298~K, in agreement with experimental observations. Strategies including cation doping and particle size reduction can be considered to flatten the potential energy landscape of the $\gamma$ phase and improve Ca mobility as a result. Our computational findings establish a theoretical rationale for the experimentally observed low capacity in CaV$_2$O$_4$, highlight the interplay between thermodynamic stability and ionic transport, and provide design strategies that can enable the practical use of CaV$_2$O$_4$ as a CB cathode.
\end{abstract}

%%%%%% Main Text %%%%%%

\section{Introduction}
\noindent Lithium-ion batteries have enabled  the portable electronics revolution and are now central to electric mobility and grid-scale storage, but the technology is approaching intrinsic limitations in both energy density and resource availability.\cite{whittingham2014ultimate, larcher2015towards, nykvist2015rapidly, tarascon2010lithium, cano2018batteries} To meet the growing global demand for sustainable and cost-effective energy storage, significant attention is turning toward beyond-lithium chemistries. In this context, multivalent batteries, based on Mg, Ca, or Zn, offer an appealing alternative by combining earth abundance with the prospect of high volumetric energy densities from metal anodes.\cite{canepa2017odyssey, ponrouch2019multivalent, blanc2020scientific, muldoon2014quest, ponrouch2016towards, arroyo2019achievements} Among the multivalent chemistries, calcium is particularly attractive as it exhibits a reduction potential close to that of lithium (-$2.87$ V vs SHE), high theoretical gravimetric ($2205$ mAh/g) and volumetric ($2073$ Ah/L) capacities, and is the fifth most abundant element in the Earth’s crust.\cite{el2020exploits, muldoon2014quest} Recent breakthroughs in electrolyte design, which now permit reversible Ca plating and stripping, have transformed the feasibility of calcium batteries (CBs) from speculation to tangible opportunity.\cite{wang2018electrolyte, li2019towards, pu2020current, shyamsunder2019reversible}  

Although Ca-based chemistry offers compelling advantage, the practical realization of CBs remains constrained by several unresolved challenges. A foremost difficulty lies in the development of electrolytes that are simultaneously stable against the highly reducing Ca metal anode and the oxidizing potentials required for high-voltage cathodes.\cite{arroyo2019achievements, monti2019multivalent} Equally critical is the identification of positive electrode (cathode) materials that are not only thermodynamically stable but also capable of reversible Ca intercalation at relevant voltages and capacities.\cite{gummow2018calcium, rong2015materials} Moreover, the intrinsically poor Ca\textsuperscript{2+} transport in most solid frameworks leads to prohibitively high migration barriers ($E_m$), limiting rate performance and cycle life. Together, these challenges highlight the complexity of advancing CB technology from conceptual promise to practical deployment.

Significant effort has been devoted to identifying viable Ca-intercalation cathodes, guided by both computational and experimental studies.\cite{tekliye2022exploration, tekliye2024fluoride,  lu2021searching, kim2020high, jeon2020reversible, black2022elucidation, tekliye2026geometry} For instance, a high-throughput screening of ternary transition-metal oxides and chalcogenides by Lu et al. identified post-spinel CaV$_2$O$_4$ and layered CaNb$_2$O$_4$ as particularly promising candidates.\cite{lu2021searching} Building on this theoretical prediction, Black et al. provided experimental validation of the redox electrochemical activity in post-spinel CaV$_2$O$_4$, while complementary density functional theory (DFT\cite{hohenberg1964inhomogeneous, kohn1965self}) calculations by the authors quantified the Ca$^{2+} E_m$ within the lattice.\cite{black2022elucidation} Nevertheless, Black et al.'s work indicated only partial calcium extraction, with capacities restricted to $\sim 65$~mAh~g$^{-1}$ at 298~K and $\sim 155$~mAh~g$^{-1}$ at 323~K, and full Ca deintercalation not achieved even at 323~K.\cite{black2022elucidation} Thus, experimental data in CaV$_2$O$_4$ indicates that poor Ca$^{2+}$ mobility, alongside possible side reactions with the electrolyte, may play a role in restricting the electrochemical capacity, as indicated by the strong temperature-dependence in the extent of Ca that is extracted. 

Additionally, the evolution of the voltage profile during cycling of CaV$_2$O$_4$,\cite{black2022elucidation} as characterized by the emergence of distinct plateaus and a decrease in polarization, suggests a possible dynamic restructuring of the host lattice or the onset of phase separation. Although operando X-ray diffraction (XRD) has confirmed the extraction of $\sim$~0.3~mol of Ca from CaV$_2$O$_4$ and the formation of oxidized phases, the precise interplay between the thermodynamic stability of the oxidized phases and any kinetic barriers associated with them is not fully understood. Thus, it is an open question whether the current electrochemical inability to reach the full theoretical capacity in CaV$_2$O$_4$ is a thermodynamic impossibility or a kinetic trap. Understanding these factors is therefore critical to fully evaluate the potential of CaV$_2$O$_4$ as a practical CB cathode, motivating the systematic investigation of our work.

Here, we present a comprehensive computational investigation of Ca intercalation in Ca$_x$V$_2$O$_4$ ($0\leq x \leq 1$) that integrates DFT calculations, the cluster expansion (CE\cite{sanchez1984generalized}) formalism, grand canonical Monte Carlo (GCMC) simulations, and nudged elastic band (NEB\cite{henkelman2000improved,sheppard2008optimization}) computations to elucidate both the thermodynamic phase stability and the Ca$^{2+}$ transport kinetics across the full Ca compositional range. We construct the 0 K convex hull and a finite temperature phase diagram to identify stable single phases and possible two phase equilibria at different temperatures, and compute the corresponding voltage profile(s). In addition, $E_m$ calculations are performed for all thermodynamically accessible single phase regions to pinpoint compositions that can pose kinetic limitations associated with Ca diffusion. Our combined thermodynamic and kinetic analysis reveals that Ca extraction is limited by the interplay between persistent phase separation during (de)intercalation and composition-dependent Ca$^{2+}$ mobility variations. Specifically, the $\delta$ ($x \sim 0.67$) and $\gamma$ ($x \sim 0.5$) two-phase region extends over a broad temperature range that encompasses electrochemical operating conditions with Ca$^{2+}$ mobility severely impeded in the $\gamma$ phase due to the strong Ca-vacancy ordering, which acts as the kinetic bottleneck for full reversible Ca (de)intercalation from CaV$_2$O$_4$. By establishing explicit links between phase behavior and ionic transport, our study provides a quantitative framework for rationalizing capacity limitations in Ca based intercalation hosts, such as CaV$_2$O$_4$, and for guiding the design of improved Ca-cathodes.

%\newpage
\section{Methods}
\subsection{First principles calculations}
\noindent All spin-polarized geometry relaxation calculations were performed using DFT as implemented in the Vienna ab initio simulation package\cite{kresse1993ab, kresse1996efficient} with the Hubbard \textit{U}-corrected strongly constrained and appropriately normed (i.e., SCAN+\textit{U}\cite{sun2015strongly, dudarev1998electron, anisimov1991band, gautam2018evaluating, long2020evaluating,tekliye2024accuracy, swathilakshmi2023performance}) exchange-correlation functional. A \textit{U} correction of 1.0 eV, as obtained from our previous work, was applied to the 3\textit{d} electrons of V to reduce the spurious self interactions.\cite{gautam2018evaluating, long2020evaluating} We used the frozen-core projector augmented wave (PAW\cite{kresse1999ultrasoft, blochl1994improved}) potentials (see \textbf{Table~S1} of the supporting information --SI, for the specific potentials used) and expanded the plane wave basis up to a kinetic energy cutoff of 520~eV. We sampled the irreducible Brillouin zone with $\Gamma$-centered Monkhorst-Pack\cite{monkhorst1976special} $k$-points grids of a minimum density of 48 $k$-points per \AA{}. Geometric relaxations of cell volume, shape, and ionic positions were carried out without symmetry constraints, with convergence criteria of $10^{-5}$ eV and $\lvert 0.03 \rvert$ eV/\AA{} for total energy and atomic forces, respectively. We obtained the starting structure of CaV$_2$O$_4$ from the inorganic crystal structure database.\cite{hellenbrandt2004inorganic}

\subsection{Cluster expansion formalism}
\noindent To obtain the temperature–composition phase diagram, we performed GCMC simulations using a CE Hamiltonian. In the CE formalism, the total energy of a structure is expressed as a function of the site occupation variables, $\sigma_i$, written as a summation over effective cluster interactions (ECIs) of point, pair, triplet, and higher-order clusters, as given in Eq.~\ref{eq:1}: 

\begin{equation}
E({\sigma}) = V_0 + \sum_{\alpha} m_{\alpha} V_{\alpha} \langle \Phi_{\alpha}({\sigma}) \rangle_{\alpha},
\label{eq:1}
\end{equation}

where ${\sigma} = (\sigma_1, \sigma_2, \ldots)$ is the configuration vector with $\sigma_i = +1$ if site $i$ is occupied by Ca and $-1$ if it is vacant. The sum runs over symmetrically distinct clusters $\alpha$ (points, pairs, triplets, \ldots), $m_{\alpha}$ is the multiplicity of cluster $\alpha$ per primitive cell, and $V_{\alpha}$ is the corresponding ECI. $\langle \Phi_{\alpha}({\sigma}) \rangle_{\alpha} = \langle \prod_{i \in \alpha} \sigma_i \rangle_{\alpha}$ is the cluster function averaged over all symmetrically equivalent clusters of type $\alpha$ in a given configuration. $V_0$ represents a constant ECI term, also referred to as the ECI of the empty cluster. The symmetrically unique ECIs were systematically fitted to a set of training structures and their corresponding DFT-calculated formation energies. Note that the DFT formation energy of a given configuration was computed from the DFT total energy of that configuration referenced to the DFT calculated energies of end-member compositions, namely, V$_2$O$_4$ ($x=0$) and CaV$_2$O$_4$ ($x=1$). 
    
To construct the training set for the CE, we enumerated unique Ca-vacancy configurations at different $x$ in Ca$_x$V$_2$O$_4$ using $1\times1\times1$, $1\times1\times2$, and $1\times1\times3$ supercells of the conventional cell that contains four Ca${_x}$V$_2$O$_4$ formula units. We used the \texttt{pymatgen} package\cite{ong2013python} to enumerate the symmetrically distinct Ca-vacancy configurations at a given $x$ and ranked them based on their electrostatic energies via the Ewald energy summation technique.\cite{ewald1921berechnung} We considered all enumerated structures across 15 distinct Ca-compositions, namely, $x = $ 0.083, 0.125, 0.167, 0.250, 0.333, 0.375, 0.417, 0.500, 0.583, 0.625, 0.667, 0.750, 0.833, 0.875, 0.917, besides the end-member compositions ($x$ = 0 and 1), resulting in a total dataset of 262 DFT-calculated formation energies across all $x$ considered. For constructing the CE, we utilized the clusters approach to statistical mechanics (\texttt{CASM}\cite{puchala2023casm, van2018first, puchala2013thermodynamics}) package, with ECIs determined using least absolute shrinkage and selection operator regression to obtain a sparse and accurate model. We evaluated the accuracy of the CE fit using the root-mean-square error (RMSE) score across all configurations, while the predictive ability of the fit was assessed using the weighted cross-validation (WCV) score.  

\subsection{Monte Carlo simulations}
\noindent All GCMC simulations were performed using the \texttt{CASM} package and our CE Hamiltonian, to obtain the temperature-composition phase diagram of Ca$_x$V$_2$O$_4$. The GCMC simulations employed a $12 \times 12 \times 12$ supercell of the conventional CaV$_2$O$_4$ cell, containing 6,912 Ca/vacancy sites. At each Ca chemical potential ($\mu$) and temperature ($T$), GCMC simulations were performed using the Metropolis algorithm\cite{metropolis1953equation} with a minimum of 10,000 and a maximum of 100,000 Monte Carlo passes. The equilibration/sampling split among the Monte Carlo passes was determined automatically using the statistical equilibration criterion of van de Walle and Asta,{\cite{van2002self}} which identifies and discards the initial equilibration portion of the Monte Carlo trajectory before ensemble averaging. Subsequently, the sampling of Monte Carlo passes continued until the requested statistical precision of 0.25 meV/f.u. on the calculated energy at a 95\% confidence level was achieved, or until the maximum number of passes was reached. We scanned from $T = 5$~K to 1005~K in steps of $\Delta T = 5$ K. At every $T$, we scanned the $\mu$ forward (i.e., increasing $\mu$) and backward (decreasing $\mu$) across four concentration regions, spanning a range of $-7.0 \leq \mu \leq 7.0$ eV/atom, with a step size of $\Delta\mu$~=~0.01~eV/atom. The four concentration ranges covered during every $\mu$ scan are $-7.0 \leq \mu \leq -3.7$, $-3.7 \leq \mu \leq 0.0$, $0.0 \leq \mu \leq 3.0$, and $3.0 \leq \mu \leq 7.0$, where $\mu$ values of -7.0, -3.7, 0.0, 3.0, and 7.0~eV/atom correspond to the five ground state structures at $x = 0, 0.25, 0.50, 0.67,$ and 1, respectively, on the 0~K convex hull.

Additionally, we used canonical Monte Carlo (CMC) based simulated annealing to verify the authenticity of the ground state configurations predicted by the CE and to ensure that all ground states at 0~K have been captured accurately. Specifically, we performed CMC simulations consisting of a minimum of 100 passes per $T$ from 2005~K to 5~K in steps of 100~K at different $x$ and under varying supercell sizes for each $x$ (for example, from 3$\times$1$\times$2 to 3$\times$3$\times$3 at $x =$0.83). During the CMC simulations, if a new configuration was predicted that was not present already in the training data, we calculated its energy using DFT and subsequently added the configuration to the training set. We repeated this process iteratively until no further ground states emerged. All CMC simulations were also performed using the \texttt{CASM} package.
    
We determined the Ca intercalation voltage ($V$, versus Ca/Ca$^{2+}$) at all $T$ directly from the $\mu$ obtained from GCMC simulations, using Equation~\ref{eq:volt_pr}:
    \begin{equation}
    V(x, T) = -\frac{\mu(x, T)}{z} + \Delta V
    \label{eq:volt_pr}
    \end{equation}
where $z=2$ is the valence of $\text{Ca}^{2+}$. $\Delta V$ is a constant shift applied to align the average calculated voltage across the entire $0 \leq x \leq 1$ range at any $T$ with the reference average intercalation voltage of 2.73~V obtained from 0~K DFT calculations. To construct the equilibrium voltage-composition ($V$-$x$) profiles as well as the temperature-composition phase diagram, the $x$ was determined at each $\mu$ by resolving the numerical hystersis within GCMC simulations through the selection of the phase that minimizes the grand canonical potential (i.e., via the free energy integration technique, see \textbf{Section~S3} of SI for details.\cite{hinuma2007phase, hinuma2008temperature, van2000phase, van2010linking}) 
    
\subsection{Migration barriers}
\noindent We performed DFT-based NEB calculations to evaluate the Ca$^{2+}~E_m$ in Ca$_x$V$_2$O$_4$ at select stable single phase compositions identified from the calculated phase diagram. Due to the trade-off between computational cost and accuracy, we used the generalized gradient approximation (GGA\cite{perdew1996generalized}) exchange–correlation functional instead of SCAN/SCAN+$U$ for all NEB calculations, as GGA is known to obtain qualitative trends in $E_m$ precisely.\cite{devi2022effect} For the endpoint configurations in an NEB calculation, we converged the structures till the atomic forces fell below $|0.03|$~eV/$\text{\AA}$. Subsequently, we initialized the minimum energy path by linear interpolation of lattice vectors and atomic positions between the endpoint configurations, generating seven intermediate images with a spring constant of 5 eV/\AA{}$^2$ applied between the images. We used $\Gamma$-centered Monkhorst-Pack $k$-point meshes containing at least 32 subdivisions along each unit reciprocal lattice vector to sample the irreducible Brillouin zone of both the endpoint and the image configurations. We considered our NEB calculations converged when the perpendicular component of the band force fell below $|0.05|$ eV/$\text{\AA}$. For compositions where DFT-based NEB calculations were not feasible, we used machine learning (ML) models to get qualitative estimates (see \textbf{Section~S7} for methodological details).

\section{Results}
\subsection{Structure of $\text{Ca}_x\text{V}_2\text{O}_4$}
\begin{figure}[h!]
\centering
\includegraphics[width=\textwidth]{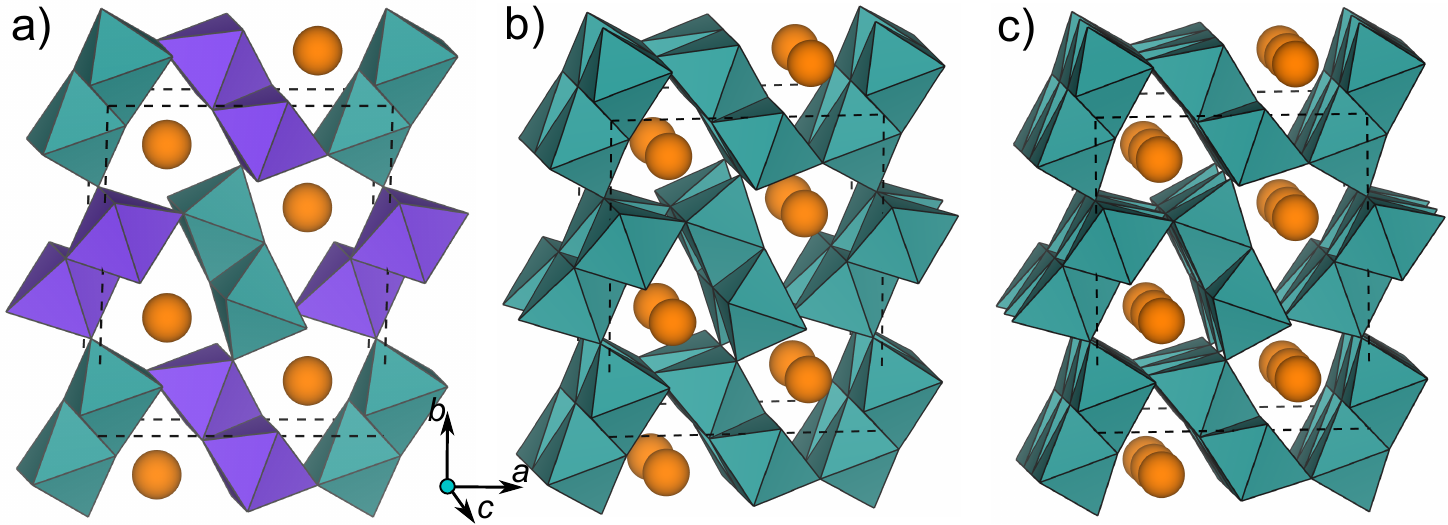}
\caption{Crystal structure of Ca$_x$V$_2$O$_4$. Teal and purple (in panel a) polyhedra denote the VO$_6$ framework around V1 and V2 sites, respectively, while orange spheres denote the Ca sites. Dashed lines outline the simulation cell. (a) Conventional cell, (b) $1\times1\times2$, and (c) $1\times1\times3$ supercells.}
\label{fig:str}
\end{figure}

\noindent Post-spinel CaV$_2$O$_4$ adopts the orthorhombic CaFe$_2$O$_4$-type structure (space group \textit{Pnma}) comprising two crystallographically distinct V octahedral sites, namely, V1 and V2 (teal and purple octahedra in \textbf{Figure~\ref{fig:str}a}).\cite{niazi2009single} The distorted VO$_6$ octahedra share both edges and corners, with Ca occupying eightfold-coordinated sites in the voids created. Ca transport within the lattice is expected to occur predominantly along the $c$-axis tunnels.\cite{lu2021searching,black2022elucidation} To construct the CE model, we use the conventional cell (\textbf{Figure~{\ref{fig:str}}a}) as the primitive lattice, as well as the larger $1\times1\times2$ (\textbf{Figure~{\ref{fig:str}}b}) and $1\times1\times3$ (\textbf{Figure~{\ref{fig:str}}c}) supercells to enumerate Ca–vacancy orderings. We use these supercell dimensions to capture longer-range interactions along the Ca-diffusion direction as well as to capture different possible values of $x$ that cannot be described within the conventional cell. 

\subsection{Convex hull and cluster expansion}
\begin{figure}[h!]
\centering
\includegraphics[width=\textwidth]{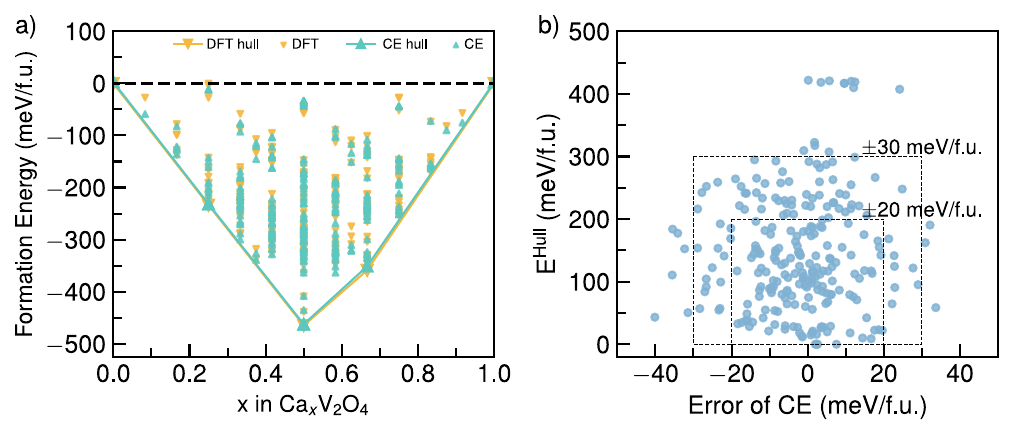}
\caption{(a) Convex hull of Ca$_x$V$_2$O$_4$ showing formation energies (meV/f.u.) vs.\ Ca content, $x$. Orange triangles denote DFT-computed energies and teal triangles the CE-computed values. Solid lines mark the $T=0$~K ground states, as estimated by DFT (orange) and CE (teal). (b) Quantifying the extent of CE error made across different configurations based on their DFT-computed energy above the convex hull (E$^{\mathrm{Hull}}$).}
\label{fig:convh}
\end{figure}

\noindent \textbf{Figure~\ref{fig:convh}a} presents the 0 K convex hull of Ca$_{x}$V$_2$O$_4$, computed using DFT (orange line) and the fitted CE Hamiltonian (teal line). The formation energies are referenced to the fully intercalated (CaV$_2$O$_4$) and deintercalated (V$_2$O$_4$) end members. Individual datapoints are shown as orange (DFT) and teal (CE) triangles, with larger symbols denoting structures that lie on the convex hull and are therefore thermodynamically stable ground states at 0~K. Notably, three intermediate compositions at $x = 0.25$, $0.5$, and $0.67$ appear as stable ground states on the 0 K convex hull, indicating the formation of distinct Ca-vacancy orderings across the Ca concentration range. Additionally, our CMC simulations indicated that no other ground state configurations appear at other $x$ near 0~K, verifying that we have a total of five ground state configurations in the Ca$_x$V$_2$O$_4$ system including the end members. The ground state structures at intermediate $x$ (0~K) are shown in \textbf{Figure~S1} of the SI.

The accuracy of the CE as a function of thermodynamic (in)stability is illustrated in \textbf{Figure~\ref{fig:convh}b}, which correlates the prediction error (represented by blue circles) with the energy above the 0 K convex hull (E$^{\text{Hull}}$) for each configuration. A robust CE model is expected to exhibit enhanced precision for structures in close proximity to the convex hull (i.e., low E$^{\text{Hull}}$ values) to ensure reliable phase stability predictions. Consistent with this requirement, the current model demonstrates a marked concentration of low-error configurations as E$^{\text{Hull}}$ approaches zero. Specifically, the majority of structures with E$^{\text{Hull}} \leq$~200~meV/f.u. of the hull fall within an error margin of $\pm$20 meV/f.u., as indicated by the inner dashed box, while nearly all configurations relevant to phase stability under electrochemical operating conditions (i.e., E$^{\text{Hull}} \leq$300~meV/f.u.) are captured within a $\pm$30 meV/f.u. threshold (outer dashed box). 

Overall, our CE model achieves an RMSE of $13.4$ meV/f.u., accurately reproduces all DFT-predicted ground states, and exhibits good transferability with a leave-one-out WCV of $15.7$ meV/f.u., confirming the model’s reliability in predicting energies of configurations not included in its fit. While the overall CE fit is excellent, a small subset of data points in the E$^{\text{Hull}} \sim$50–200~meV/f.u. range do exhibit relatively large deviations, as indicated by the blue circles lying outside the dashed boxes in \textbf{Figure~\ref{fig:convh}b}. These outliers likely arise from complex local environments or significant structural relaxations during DFT that are inherently challenging to capture within a truncated CE framework.

\begin{figure}[h!]
\centering
\includegraphics[width=\textwidth]{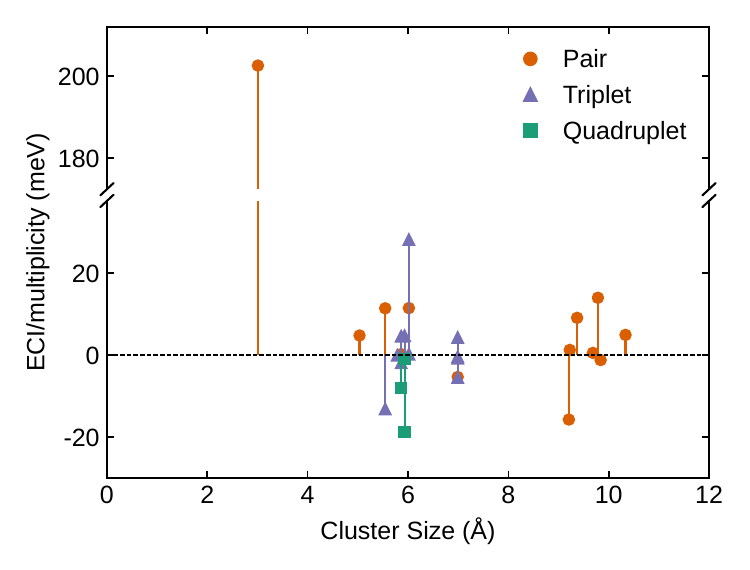}
\caption{Plot of ECIs, normalized by multiplicity, as a function of cluster size for $\text{Ca}_x\text{V}_2\text{O}_4$. Orange circles, purple triangles, and teal squares denote pair, triplet, and quadruplet interactions, respectively. Note the break in the $y$-axis done to accommodate the dominant pair term within the plot.}
\label{fig:eci}
\end{figure}

\textbf{Figure~\ref{fig:eci}} illustrates the 29 ECIs that constitute our CE model, as obtained by our fit against the training set of 262 DFT configurations. The details of all clusters that exhibit a non-zero ECI in our fit, including the empty and point terms, are summarized in \textbf{Table~S2} of the SI. We normalize the ECIs by multiplicity in \textbf{Figure~{\ref{fig:eci}}} and plot the values against the corresponding size of the cluster to highlight the physical hierarchy of the CE. Note that clusters are categorized by their maximum distance, i.e., the size of a triplet is the distance between its farthest sites. The ECIs for pairs, triplets, and quadruplets are indicated by orange circles, purple triangles, and teal squares, respectively, with the empty and point ECI values listed in \textbf{Table~S2}. For pair interactions, positive ECIs indicate repulsive interactions (i.e., Ca-vacancy pairs are favored) while negative ECIs signify attractive interactions (i.e., Ca-Ca and vacancy-vacancy pairs are favored). 

Our CE exhibits a well-converged and clear physical hierarchy, characterized by the rapid decay of ECIs with both increasing cluster size and increasing cluster complexity. For example, the most dominant Ca-Ca interaction in the Ca$_x$V$_2$O$_4$ system is the nearest repulsive pair interaction, at $\sim$3~\AA{} with a value of $\sim$202~meV, indicating the strong and short-ranged electrostatic interaction that governs Ca ordering in the structure. In other words, Ca strongly prefers vacancies as nearest neighbors within the V$_2$O$_4$ lattice, which is reflected in the diamond-cubic-like arrangement of Ca atoms and vacancies within the Ca sub-lattice at the $x = 0.5$ ground state configuration (see \textbf{Figures~S1b} and \textbf{\ref{fig:convh}a}). As the size of the pair clusters increases, the ECI magnitudes generally decrease and fluctuate around zero, with values in the range of $\pm$15~meV over the cluster-size range of $\sim$5--10~\AA{}. In terms of cluster complexity, the dominant triplet (at $\sim$6~\AA{}) and quadruplet (also at $\sim$6~\AA{}) interactions remain significantly smaller than the dominant pair interaction, with ECIs of approximately 28 and $-19$~meV, respectively. Overall, the ECIs indicate the dominant role played by pair interactions (with lower complexity) compared to triplet and quadruplet (higher complexity) interactions. 
   
\subsection{Temperature-composition phase diagram}
\begin{figure}[h!]
\centering
\includegraphics[width=\textwidth]{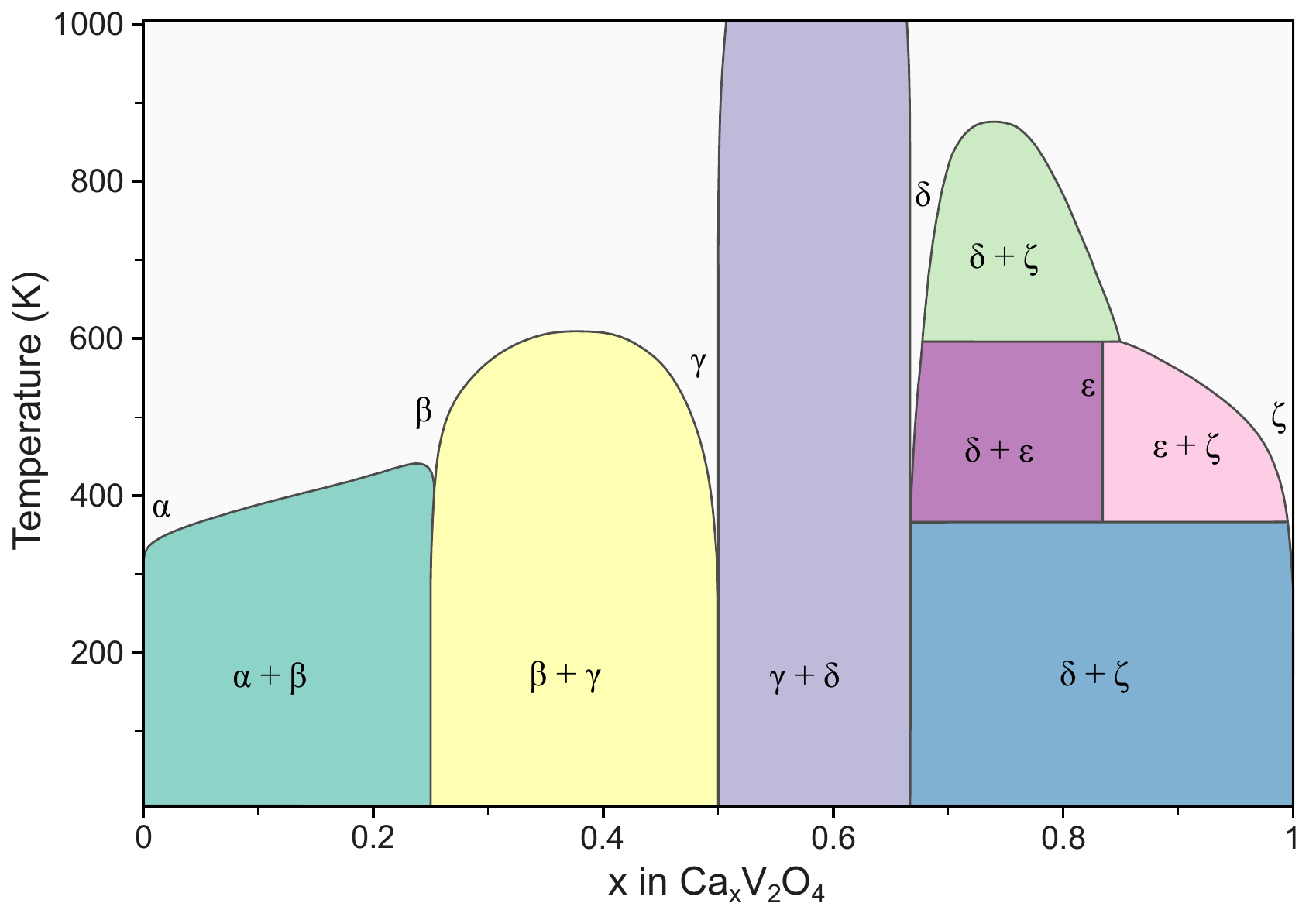}
\caption{Computed temperature-composition ($T-x$) phase diagram for the $\text{Ca}_x\text{V}_2\text{O}_4$ system. The colored regions represent the co-existence of two-phases that are separated by the single-phase regions ($\alpha$ through $\zeta$).}
\label{fig:phasediagram}
\end{figure}

\noindent Elucidating the temperature-dependent phase behavior of the CaV$_2$O$_4$ system, \textbf{Figure~{\ref{fig:phasediagram}}} illustrates the calculated temperature-composition phase diagram, as derived from GCMC simulations utilizing large supercells. The associated occupancy of Ca atoms across the four Ca-sites in the CaV$_2$O$_4$ lattice, as a function of $x$ and $T$ is compiled in \textbf{Figure~S2}. The computed phase diagram in \textbf{Figure~{\ref{fig:phasediagram}}} reveals a complex thermodynamic landscape comprising six distinct single-phase regions, denoted as $\alpha$ through $\zeta$. At temperatures below 298~K, five distinct single-phase regions exist, arising from the corresponding 0~K ground state configurations (\textbf{Figure~\ref{fig:convh}a}), namely, $\alpha, \beta, \gamma, \delta,$ and $\zeta$, corresponding to $x = 0, 0.25, 0.5, 0.67,$ and $1$, respectively. Notably, the $\varepsilon$-phase (situated at $x \sim 0.83$) is unique in being stabilized at higher temperatures (via configurational entropic contributions) and does not have a corresponding ground state configuration at 0~K. The $\varepsilon$ phases emerges only at an elevated $T \sim$~370~K and persists until 590~K, with the associated invariant reactions at 370~K and 590~K being of the eutectic- and peritectic-types, respectively. A representative snapshot of the structure of the $\varepsilon$ phase at 370~K is shown in \textbf{Figure~S3} with the corresponding crystallographic information file provided in our \href{https://github.com/sai-mat-group/CaV2O4-phase-diagram}{GitHub} repository. 

The two-phase regions in \textbf{Figure~\ref{fig:phasediagram}} are shown as colored domes, with gray lines marking their boundaries. Specifically, the $\alpha + \beta$ two-phase region exists between $0 \leq x \leq 0.25$ (teal dome), with a narrow solid-solution (white region) beginning at $T \sim$~305~K near $x = 0$, signifying negligible solubility of Ca in the V$_2$O$_4$ structure at 298~K. The $\alpha + \beta$ two-phase region remains until $\sim$440~K at $x \sim 0.25$, with the system transitioning into a disordered solid solution phase above 440~K. Thus, Ca's solubility in V$_2$O$_4$ should increase dramatically, if electrochemical discharge into the V$_2$O$_4$ lattice is done under elevated temperatures (e.g., 323~K) compared to room temperature. Similarly, the $\beta + \gamma$ region exists between $0.25 \le x \le 0.5$ (yellow dome), with the two-phase region transitioning into a disordered solid solution at $\sim$610~K. Notably, the $\gamma + \delta$ region, spanning the composition range $0.5 \le x \le 0.67$ (light purple dome), remains stable at temperatures exceeding 1000~K (not shown in \textbf{Figure~{\ref{fig:phasediagram}}}), indicating that this two-phase region will be accessed under all electrochemical conditions irrespective of Ca intercalation into V$_2$O$_4$ or Ca removal from CaV$_2$O$_4$. Also, the $\delta$ phase exhibits fairly negligible non-stoichiometry even at elevated temperatures (up to $\sim$~590~K), further contributing to the persistence of the $\gamma + \delta$ two-phase region.

The composition range between $0.67 \le x \le 1.0$ displays a more complex set of transitions compared to $x < 0.67$. At low temperatures, the $\delta + \zeta$ two-phase region (blue) exists up to 370 K, with negligible solubility of vacancies in the CaV$_2$O$_4$ lattice (i.e., $\zeta$ phase). Above 370~K, the emergence of the $\varepsilon$ phase via a eutectic-type reaction causes the formation of two distinct two-phase regions, namely, $\delta + \varepsilon$ (purple) and $\varepsilon + \zeta$ (pink), with a corresponding increase in vacancy solubility in the $\zeta$ phase. As $T$ increases further, the $\varepsilon$ phase disappears via a peritectic-type reaction at 590 K, resulting in the formation of a single two-phase region, $\delta + \zeta$ (green dome), which persists till $\sim$855~K. The extent of off-stoichiometry in both the $\delta$ and $\zeta$ phases reaches an inflection point that coincides with the peritectic-type reaction at 590~K and continues to increase rapidly with further increase in $T$. Eventually, the $0.67 \le x \le 1.0$ composition range exhibits a single disordered solid solution at $T >$~855~K. Given that both the solubility of Ca in V$_2$O$_4$ ($\alpha$ phase) and of vacancy in CaV$_2$O$_4$ ($\zeta$ phase) are negligible at 298~K and exhibit an increase at higher temperatures, the Ca$_x$V$_2$O$_4$ cathode may exhibit better electrochemical cycling (i.e., with lower polarization) at elevated temperatures, in qualitative agreement with experimental observations so far.\cite{black2022elucidation} 

The occupancy of Ca atoms among the four distinct Ca sites in the V$_2$O$_4$ lattice (\textbf{Figure~{\ref{fig:str}}a}) shows strong non-monotonic behavior and temperature dependence as well (see \textbf{Figure~S2}). However, the occupancies of all Ca sites in the V$_2$O$_4$ lattice converge to a value of 0.5 (or half occupancy) as $x$ approaches 0.5 ($\gamma$ phase) at all temperatures, indicating the strong nature of the Ca-vacancy ordering in the $\gamma$ phase (\textbf{Figure~S1b}) that is driven by strong electrostatic attraction between Ca atoms and neighboring vacancies (\textbf{Figure~{\ref{fig:eci}}}). In comparison to $x \leq 0.5$, Ca occupancies exhibit fairly linear trends across all four Ca sites in the $x > 0.5$ composition range (\textbf{Figure~S2}), across all temperatures, with the occupancy of all four sites being equal in the $\varepsilon$ single phase at 475~K. 

\subsection{Voltage profiles}
\begin{figure}[h!]
\centering
\includegraphics[width=\textwidth]{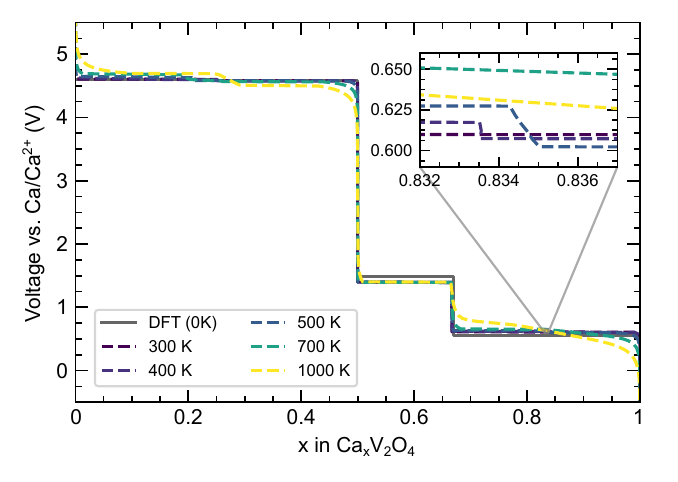}
\caption{Computed voltage profiles for Ca intercalation in $\text{Ca}_x\text{V}_2\text{O}_4$ as a function of $x$. The finite-temperature voltage profiles (dashed lines) are derived from GCMC simulations and are color-coded by temperature. The solid gray line represents the 0~K DFT-computed voltage profile. The inset highlights the voltage step in the $x \sim 0.83$ region, corresponding to the $\varepsilon$-phase.}
\label{fig:voltage}
\end{figure}

\noindent \textbf{Figure~\ref{fig:voltage}} presents the GCMC-computed voltage profiles for Ca intercalation in $\text{Ca}_x\text{V}_2\text{O}_4$, color-coded by temperature from dark purple (300~K) to bright yellow (1000~K), and are compared against the 0~K DFT-computed voltage profile (solid gray). The inset in \textbf{Figure~\ref{fig:voltage}} focuses on the narrow composition range of 0.832~$< x <$~0.836, which is adjacent to the $\varepsilon$ phase. Notably, the 0~K DFT and the 300~K GCMC voltage profiles exhibit three distinct voltage steps at $x = 0.25$ ($\beta$), $x = 0.5$ ($\gamma$), and $x = 0.67$ ($\delta$), corresponding to the corresponding stable single-phase regions (\textbf{Figure~{\ref{fig:phasediagram}}}). Consequently, four well-defined voltage plateaus are observed, representing the two-phase coexistence regions: $\alpha + \beta$ ($\sim$ 4.60~V vs. Ca/Ca$^{2+}$), $\beta + \gamma$ ($\sim$ 4.59~V), $\gamma + \delta$ ($\sim$ 1.40~V), and $\delta + \zeta$ ($\sim$ 0.60~V). Thus, the $\alpha + \beta$ and/or the $\beta + \gamma$ two-phase regions need to be accessed electrochemically for Ca$_x$V$_2$O$_4$ to exhibit high (de)intercalation voltage (and hence high energy density) as a cathode. While a one-one comparison with the experimental voltage profile is not possible due to significant polarisation across the charge-discharge profiles and differences in the voltage plateaus (or steps) across the charging and discharging steps,\cite{black2022elucidation} we do note that our overall calculated average voltage across the entire Ca composition range (2.73~V vs. Ca/Ca$^{2+}$) is lower than the experimental average voltage of $\sim$~3.55~V across a 0.6 mol Ca extraction from CaV$_2$O$_4$. However, side reactions and overpotentials during Ca extraction from CaV$_2$O$_4$ may also be contributing to the observed voltages experimentally, causing the disagreement between computations and experiments.

Notably, at 400 K and 500 K, the computed voltage profile exhibits an additional (small) step corresponding to the emergence of the $\varepsilon$ ($x \sim 0.83$) at 370 K, as highlighted in the inset of \textbf{Figure~\ref{fig:voltage}}. Note that this voltage step occurs at a composition that is marginally offset from the $\varepsilon$ composition (by $\sim$~0.001-0.002), possibly due to sampling issues in our GCMC and the associated free energy integration. Nevertheless, our computational finding of the presence of the $\varepsilon$ phase is in agreement with the experimental electrochemical profiles, which show subtle slope changes and capacity plateaus within the $x \sim 0.80$ to $0.85$ range in $\text{Ca}_{x}\text{V}_2\text{O}_4$,\cite{black2022elucidation} suggesting a possible (metastable) formation of the $\varepsilon$ phase below 370~K. As the temperature increases to 1000 K (bright yellow), the voltage profile evolves from sharp, discrete steps into a smoother series of plateaus, signifying the existing of the disordered single phase regions at $x <$~0.5 and $x >$~0.67 (\textbf{Figure~{\ref{fig:phasediagram}}}). Notably, the voltage plateau corresponding to the $\gamma + \delta$ region ($0.5 \le x \le 0.67$) remains even at 1000~K, consistent with our calculated phase diagram, and in qualitative agreement with the experimental charging curves at 323~K.\cite{black2022elucidation}

\subsection{Ca$^{2+}$ migration barriers}
\begin{figure}[h!]
\centering
\includegraphics[width=\textwidth]{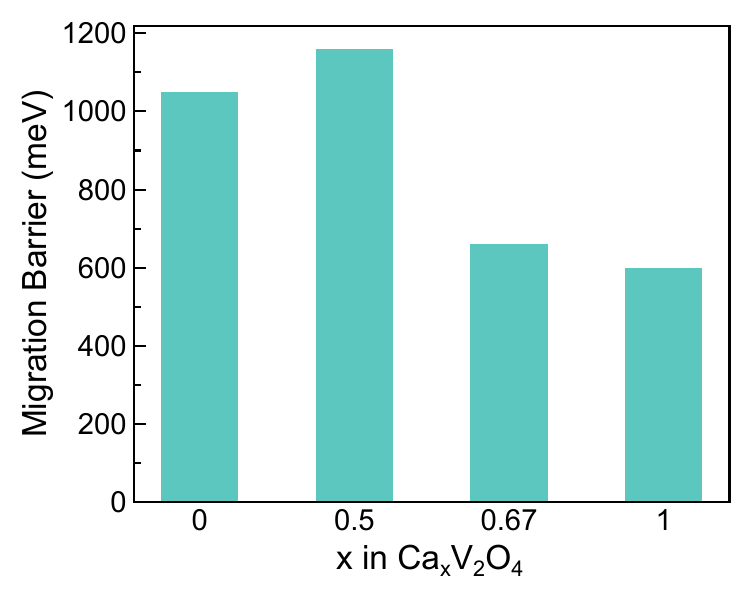}
\caption{DFT-based NEB calculated Ca$^{2+}$ $E_m$ in Ca$_x$V$_2$O$_4$ at different $x$. Results for the endpoints ($x=0$ and $x=1$) are obtained Ref.~\citenum{black2022elucidation}, while the remaining $E_m$ are calculated in this work. \textbf{Figure~S5} compiles the $E_m$ estimates for the different ground states in Ca$_x$V$_2$O$_4$ obtained using machine learned models.}
\label{fig:7barrier}
\end{figure}

\noindent Given the dominant two phase region across the $0.5 \le x \le 0.67$ range (\textbf{Figure~{\ref{fig:phasediagram}}}), any limitation on Ca-mobility within the $\gamma$ or the $\delta$ phases can adversely affect the amount of Ca that can be reversibly (de)intercalated into the Ca$_x$V$_2$O$_4$ system. Hence, we calculate the $E_m$ for Ca-motion using DFT-based NEB in different ground state ordered structures of Ca$_x$V$_2$O$_4$ and present the data for $x = 0, 0.5, 0.67$, and 1 in \textbf{Figure \ref{fig:7barrier}}. The endpoint values ($x = 0$ and $x = 1$) are based on the GGA functional and obtained from the previous work of Black et al.,\cite{black2022elucidation} while we have computed the $E_m$ for $x = 0.5$ and 0.67 in this work and have included the associated minimum energy pathways in \textbf{Figure~S4}. Due to the supercell size requirements for calculating $E_m$ at $x = 0.25$ ($\beta$) and $x = 0.83$ ($\varepsilon$), we did not perform DFT-based NEB calculations for these phases and instead used foundational machine learned interatomic potentials (MLIPs\cite{batatia2025design, rhodes2025orb}, without any fine-tuning) and a transfer learned (TL) model\cite{devi2026leveraging} to obtain $E_m$ estimates for these phases, similar to our previous work\cite{tekliye2026geometry} (see \textbf{Figure~S5}). MLIP and TL estimates of $E_m$ for the $x = 0, 0.5, 0.67$, and 1 ground states are also provided in \textbf{Figure~S5} for comparison.

\subsection{Voltage profiles}

The $E_m$ exhibit a strong, non-monotonic dependence on the calcium content, with the peak (minimum) in $E_m$ at $x = 0.5$ ($x = 1$). In the high-Ca limit (i.e., $x = 1$, $\zeta$ phase), the $E_m$ is low ($\sim$~600~meV), suggesting favorable Ca$^{2+}$ motion under typical electrochemical conditions, as long as a sufficient number of diffusion carriers (i.e., vacancies) are present in the $\zeta$ phase. Indeed, a $E_m$ threshold of 525-650~meV has been used before to identify cathode materials with reasonable ionic mobility for various battery frameworks.\cite{rong2015materials} At $x = 0.67$ ($\delta$ phase), the $E_m$ increases marginally to $\sim$~661~meV, indicating that the formation of the $\delta$ phase during electrochemical Ca (dis)charge in Ca$_x$V$_2$O$_4$ should not act as a mobility bottleneck. 

In contrast to the $\zeta$ and $\delta$ phases, in the high-vacancy limit ($x = 0, \alpha$ phase), the $E_m$ is $\sim$~1050~meV, suggesting inherently poor Ca$^{2+}$ motion, despite the presence of a large number of vacancies. Unfortunately, the $E_m$ reaches a maximum value of $\sim$~1160~meV at $x = 0.5$ ($\gamma$ phase). Such high $E_m$ values, exceeding 1~eV in the $\gamma$ and $\alpha$ phases, indicate that Ca$^{2+}$ mobility is severely hindered in the low-to-intermediate concentration regimes ($x \le 0.5$), highlighting the kinetic bottleneck(s) associated with the reversible electrochemical cycling of Ca in the Ca$_x$V$_2$O$_4$ system. Particularly, the mobility limitation associated with the $\gamma$ phase, which will co-exist alongside the $\delta$ phase as $x$ drops below 0.67 during Ca extraction, may be the predominant reason to the limited capacity observed during electrochemical extraction of Ca from the CaV$_2$O$_4$ structure.\cite{black2022elucidation} Consequently, the accessible electrochemical capacity of CaV$_2$O$_4$ is likely limited to half of its theoretical value, unless strategies are devised to overcome the substantial kinetic bottlenecks associated with deep Ca de-intercalation from CaV$_2$O$_4$.

In terms of MLIP and TL estimates, we find that the ML predictions do not systematically reproduce the trends observed in the DFT-calculated $E_m$ (\textbf{Figure~S5}). For example, at $x = 1$ both MLIPs substantially overestimate the DFT-calculated $E_m$ (892-1093~meV vs. 600~meV), while the TL model is in better agreement (467~meV). At $x = 0.67$, all ML models do overestimate but are in better alignment with DFT-$E_m$ (690-878~meV vs. 661~meV). On the other hand, at $x = 0$ and 0.5, the ML models underestimate the DFT-$E_m$. For the $\varepsilon$ phase ($x = 0.83$) where we do not have a DFT-$E_m$, the ML-predicted $E_m$ resemble the distribution of values obtained for the $\zeta~(x = 1)$ phase (MLIPs: 921-1130~meV; TL: 636~meV). Therefore, we suggest the ML models to have also overestimated the $E_m$ at $x = 0.83$ and we expect the true DFT-$E_m$ for the $\varepsilon$ phase to exhibit a value intermediate to that of the $\zeta$ (600~meV) and $\delta$ (661~meV) phases. Hence, we do not expect the possible formation of the $\varepsilon$ phase during electrochemical cycling to be a potential bottleneck for Ca motion.

\section{Discussion}
\noindent We have performed a comprehensive first-principles investigation of the thermodynamics and Ca$^{2+}$ transport in $\mathrm{Ca}_x\mathrm{V}_2\mathrm{O}_4$ cathode to elucidate the origin of experimentally observed electrochemical capacity limitations. Ground state configurations were identified through exhaustive enumeration of configurational degrees of freedom (\textbf{Figures~{\ref{fig:str}}} and \textbf{S1}) and the resultant construction of the 0 K convex hull. A CE Hamiltonian trained on our DFT calculated Ca-vacancy configurations (\textbf{Figures~{\ref{fig:convh}}} and \textbf{\ref{fig:eci}}) was subsequently employed in GCMC simulations to map the temperature–composition phase diagram (\textbf{Figure~{\ref{fig:phasediagram}}}), thereby resolving the phase stability and the temperature-dependent voltage profiles (\textbf{Figure~{\ref{fig:voltage}}}). We computed $E_m$ for all thermodynamically accessible ordered single-phase configurations (\textbf{Figure~{\ref{fig:7barrier}}}), indicating kinetic bottlenecks associated with Ca$^{2+}$ motion alongside the formation of the $\gamma$ ($x = 0.5$) phase, which is likely the predominant reason for the observed reductions in electrochemical capacity upon Ca deintercalation from CaV$_2$O$_4$. 

Our computed phase diagram is generally aligned with the experimental observations by Black et al.\cite{black2022elucidation}. For example, our calculations predict the existence of distinct stable phases ($\alpha$ through $\zeta$, \textbf{Figure~{\ref{fig:phasediagram}}}) separated by miscibility gaps, such as the $\gamma + \delta$ gap that persists to high temperatures, and the $\delta + \zeta$ gap that can influence electrochemical operation. Operando XRD analysis by Black et al.\cite{black2022elucidation} revealed that during oxidation the Ca-stoichiometry ($x$ in Ca$_x$V$_2$O$_4$) of the oxidized phase(s) that formed remains constant even as the total Ca content in the electrode decreased, which is a likely signature of a two-phase behavior, in agreement with our predictions of miscibility gaps in the $x > 0.5$ range. Additionally, XRD refinement of Ca occupancies in the oxidised phase at 298~K indicated a Ca content of $\sim 0.69$~mol,\cite{black2022elucidation} which is in alignment with our predicted stability of the $\delta$ phase at $x = 0.67$ (\textbf{Figures~{\ref{fig:convh}}a} and \textbf{\ref{fig:phasediagram}}). 

Although quantitative comparison of calculated voltage profiles (\textbf{Figure~{\ref{fig:voltage}}}) with experimental electrochemical profiles has proven difficult, we do observe a few qualitative similarities. For example, our prediction of the formation of the $\varepsilon$ phase (\textbf{Figures~{\ref{fig:phasediagram}}} and \textbf{S3}) is in agreement with the characteristic voltage step observed near $x \sim 0.83$ in experiments,\cite{black2022elucidation} suggesting the (metastable) formation of the $\varepsilon$ phase under electrochemical charging. Additionally, we do not expect the $\varepsilon$ phase to exhibit any kinetic limitations associated with Ca removal from CaV$_2$O$_4$, based on the similarities in the ML-predicted $E_m$ values at $x = 0.83$ and $x = 1$ (\textbf{Figure~S5}). However, we note here that all ML models considered in ths work do not exhibit any qualitative agreement with the overall DFT-calculated $E_m$ trends, indicating further refinement of the models are necessary for more accurate predictions.

Our calculations indicate that the primary bottleneck that limits Ca extraction from CaV$_2$O$_4$ is the high $E_m$ for Ca$^{2+}$ motion in the $\gamma$ ($x = 0.5$) phase (\textbf{Figure~{\ref{fig:7barrier}}}). As highlighted by its large (negative) formation energy at 0~K($\sim$~-465~meV, \textbf{Figure~{\ref{fig:convh}}a}), the $\gamma$ phase exhibits a strong diamond-cubic-like ordering of Ca atoms and vacancies (\textbf{Figure~S1b}) with all four Ca sites being exactly half-occupied (\textbf{Figure~S2}), as driven by the strong electrostatic attraction between Ca atoms and neigboring vacancies (\textbf{Figure~{\ref{fig:eci}}}), which contributes to the observed high $E_m$. Ideally, Ca extraction from CaV$_2$O$_4$ can proceed until $x = 0.5$ since the two-phase $\gamma + \delta$ co-existence (\textbf{Figure~{\ref{fig:phasediagram}}}) also has the $\delta$ phase that does not exhibit any Ca-mobility limitations (\textbf{Figure~{\ref{fig:7barrier}}}). However, Ca extraction within the $\gamma + \delta$ two-phase region may be limited because of the high $E_m$ in the $\gamma$ phase, which can explain the limited extraction of $\sim$~0.3~mol of Ca from CaV$_2$O$_4$ at 298~K.\cite{black2022elucidation} Further, electrochemical measurements conducted at 323~K do enable higher extraction of Ca (up to 0.6 mol from CaV$_2$O$_4$\cite{black2022elucidation}, with possible side reactions as well contributing to the observed capacity), pointing to Ca extraction being a thermally activated process, a signature of the presence of kinetic bottlenecks in the system.

Elucidating the thermodynamic phase stability of electrode materials has long been central to rational battery design. A classic example is Li$_x$FePO$_4$ (LFP), in which control of phase behavior leads to dramatic improvements in rate capability.\cite{malik2011kinetics} Although equilibrium thermodynamics predicts a two-phase reaction separating into Li-rich and Li-poor domains in LFP,\cite{zhou2006configurational} particle size reduction enables access to a non-equilibrium solid solution during cycling, effectively bypassing the nucleation barrier and sustaining rapid electrochemical kinetics.\cite{liu2014capturing} Similar strategies have since been widely applied, and numerous studies have focused on understanding the thermodynamic phase stability of cathode materials across other battery chemistries, including sodium-ion systems.\cite{deng2020phase, wang2022phase, lee2024thermodynamics} Analogous considerations are likely crucial for optimizing the rate performance and cyclability of Ca$_x$V$_2$O$_4$. By identifying the thermodynamically stable phases and quantifying the associated kinetic barriers, this work provides a framework for exploring size-dependent effects and metastable reaction pathways as potential routes to overcome the poor kinetics observed at intermediate Ca compositions in Ca$_x$V$_2$O$_4$.

In this context, the relatively high Ca migration barrier at $x = 0.5$ is the result of an energetically unfavorable and spatially constrained diffusion landscape. One possible strategy to mitigate this limitation is to disrupt such ordering of Ca and vacancies through partial substitution at the Ca sites with electrochemically inactive (or active) cations. Similar to observations in P2-type layered sodium transition-metal oxides,\cite{xu2025understanding} where Li substitution suppresses Na-vacancy ordering and promotes a more disordered alkali-ion distribution, introducing suitable dopants in Ca$_x$V$_2$O$_4$ could break the long-range Ca--vacancy correlations, flatten the potential energy landscape, and facilitate Ca motion. Consequently, such an approach may provide a viable pathway to improve the rate capability, accessible capacity, and overall electrochemical performance of Ca$_x$V$_2$O$_4$ as a Ca cathode material.

\section{Conclusion}
\noindent We systematically investigated the thermodynamic and kinetic landscape of Ca (de)intercalation in Ca$_x$V$_2$O$_4$, a candidate cathode material for CBs, to elucidate the origins of the limited electrochemical capacity observed experimentally. Using first principles calculations combined with cluster expansion and Monte Carlo simulations, we constructed a temperature--composition phase diagram to resolve the phase stability across $x$ in Ca$_x$V$_2$O$_4$ and across temperatures, with the computed voltage profile(s) showing qualitative alignment with experiments. The 0 K convex hull and finite-temperature phase diagram revealed six distinct phases around $x = 0$ ($\alpha$), 0.25 ($\beta$), 0.5 ($\gamma$), 0.67 ($\delta$), 0.83 ($\varepsilon$), and 1.0 ($\zeta$), highlighting the complex phase behavior governing Ca (de)intercalation, with the $\varepsilon$ phase appearing over a 370-590~K temperature window via invariant reactions. Further, we demonstrated that the $\alpha$ and $\gamma$ phases (i.e., at $x \leq$~0.5), exhibit high migration barriers for Ca diffusion, which, together with the stable Ca--vacancy ordering in the $\gamma$ phase, gives rise to a significant kinetic bottleneck for full Ca extraction from CaV$_2$O$_4$. Our results establishes a fundamental basis for design strategies to improve the electrochemical performance of Ca$_x$V$_2$O$_4$, including doping, particle size reduction, and high-temperature electrochemical operations, which can eventually enable the use of Ca$_x$V$_2$O$_4$ as a cathode and the practical deployment of CBs for energy storage.

\section*{Acknowledgments}
\noindent G.S.G. acknowledges financial support from the Science and Engineering Research Board (SERB), Department of Science and Technology, Government of India, under sanction number IPA/2021/000007. D.B.T. and J.A.D. acknowledge financial assistance from the Indian Institute of Science (IISc). The authors acknowledge the computational resources provided by the Supercomputer Education and Research Centre (SERC) at IISc, and the J\"{u}lich Supercomputing Centre at Forschungszentrum J\"{u}lich, Germany, for access to the JURECA supercomputer under projects hpc-prf-emdft and hpc-prf-desal.

\section*{Data and code availability}
\noindent All computed data associated with this work are available freely to all via our \href{https://github.com/sai-mat-group/CaV2O4-phase-diagram}{GitHub} repository.

\section*{Conflicts of interest}
\noindent The authors have no conflicts of interest to declare.
%\subsection*{Funding}

%\section*{Guidelines for References}

\newpage
\bibliographystyle{unsrt}
\bibliography{achemso}

@article{whittingham2014ultimate,
  title={Ultimate limits to intercalation reactions for lithium batteries},
  author={Whittingham, M Stanley},
  journal={Chemical Reviews},
  volume={114},
  number={23},
  pages={11414--11443},
  year={2014},
  publisher={ACS Publications}
}

@article{larcher2015towards,
  title={Towards greener and more sustainable batteries for electrical energy storage},
  author={Larcher, Dominique and Tarascon, Jean-Marie},
  journal={Nature Chemistry},
  volume={7},
  number={1},
  pages={19--29},
  year={2015},
  publisher={Nature Publishing Group}
}

@article{nykvist2015rapidly,
  title={Rapidly falling costs of battery packs for electric vehicles},
  author={Nykvist, Bj{\"o}rn and Nilsson, M{\aa}ns},
  journal={Nature Climate Change},
  volume={5},
  number={4},
  pages={329--332},
  year={2015},
  publisher={Nature Publishing Group}
}

@article{tarascon2010lithium,
  title={Is lithium the new gold?},
  author={Tarascon, Jean-Marie},
  journal={Nature Chemistry},
  volume={2},
  number={6},
  pages={510--510},
  year={2010},
  publisher={Nature Publishing Group}
}

@article{cano2018batteries,
  title={Batteries and fuel cells for emerging electric vehicle markets},
  author={Cano, Zachary P and Banham, Dustin and Ye, Siyu and Hintennach, Andreas and Lu, Jun and Fowler, Michael and Chen, Zhongwei},
  journal={Nature Energy},
  volume={3},
  number={4},
  pages={279--289},
  year={2018},
  publisher={Nature Publishing Group}
}

@article{canepa2017odyssey,
  title={Odyssey of multivalent cathode materials: open questions and future challenges},
  author={Canepa, Pieremanuele and Sai Gautam, Gopalakrishnan and Hannah, Daniel C and Malik, Rahul and Liu, Miao and Gallagher, Kevin G and Persson, Kristin A and Ceder, Gerbrand},
  journal={Chemical Reviews},
  volume={117},
  number={5},
  pages={4287--4341},
  year={2017},
  publisher={ACS Publications}
}

@article{ponrouch2019multivalent,
  title={Multivalent rechargeable batteries},
  author={Ponrouch, Alexandre and Bitenc, Jan and Dominko, Robert and Lindahl, Niklas and Johansson, Patrik and Palac{\'\i}n, M Rosa},
  journal={Energy Storage Materials},
  volume={20},
  pages={253--262},
  year={2019},
  publisher={Elsevier}
}

@article{arroyo2019achievements,
  title={Achievements, challenges, and prospects of calcium batteries},
  author={Arroyo-de Dompablo, M Elena and Ponrouch, Alexandre and Johansson, Patrik and Palac{\'\i}n, M Rosa},
  journal={Chemical Reviews},
  volume={120},
  number={14},
  pages={6331--6357},
  year={2019},
  publisher={ACS Publications}
}

@article{blanc2020scientific,
  title={Scientific challenges for the implementation of Zn-ion batteries},
  author={Blanc, Lauren E and Kundu, Dipan and Nazar, Linda F},
  journal={Joule},
  volume={4},
  number={4},
  pages={771--799},
  year={2020},
  publisher={Elsevier}
}

@article{muldoon2014quest,
  title={Quest for nonaqueous multivalent secondary batteries: magnesium and beyond},
  author={Muldoon, John and Bucur, Claudiu B and Gregory, Thomas},
  journal={Chemical Reviews},
  volume={114},
  number={23},
  pages={11683--11720},
  year={2014},
  publisher={ACS Publications}
}

@article{ponrouch2016towards,
  title={Towards a calcium-based rechargeable battery},
  author={Ponrouch, Alexandre and Frontera, Carlos and Bard{\'e}, Fanny and Palac{\'\i}n, M Rosa},
  journal={Nature Materials},
  volume={15},
  number={2},
  pages={169--172},
  year={2016},
  publisher={Nature Publishing Group}
}

@article{el2020exploits,
  title={Exploits, advances and challenges benefiting beyond Li-ion battery technologies},
  author={El Kharbachi, Abdel and Zavorotynska, Olena and Latroche, M and Cuevas, Ferm{\`\i}n and Yartys, Volodymyr and Fichtner, M},
  journal={Journal of Alloys and Compounds},
  volume={817},
  pages={153261},
  year={2020},
  publisher={Elsevier}
}

@article{monti2019multivalent,
  title={Multivalent batteries—prospects for high energy density: Ca batteries},
  author={Monti, Damien and Ponrouch, Alexandre and Araujo, Rafael B and Barde, Fanny and Johansson, Patrik and Palac{\'\i}n, M Rosa},
  journal={Frontiers in Chemistry},
  volume={7},
  pages={79},
  year={2019},
  publisher={Frontiers Media SA}
}

@article{hohenberg1964inhomogeneous,
  title={Inhomogeneous electron gas},
  author={Hohenberg, Pierre and Kohn, Walter},
  journal={Physical Review},
  volume={136},
  number={3B},
  pages={B864},
  year={1964},
  publisher={APS}
}

@article{kohn1965self,
  title={Self-consistent equations including exchange and correlation effects},
  author={Kohn, Walter and Sham, Lu Jeu},
  journal={Physical Review},
  volume={140},
  number={4A},
  pages={A1133},
  year={1965},
  publisher={APS}
}

@article{lu2021searching,
  title={Searching Ternary Oxides and Chalcogenides as Positive Electrodes for Calcium Batteries},
  author={Lu, Wang and Wang, Juefan and Sai Gautam, Gopalakrishnan and Canepa, Pieremanuele},
  journal={Chemistry of Materials},
  volume={33},
  number={14},
  pages={5809--5821},
  year={2021},
  publisher={ACS Publications}
}

@article{black2022elucidation,
  title={Elucidation of the redox activity of Ca$_2$MnO$_{3.5}$ and CaV$_2$O$_4$ in calcium batteries using operando XRD: charge compensation mechanism and reversibility},
  author={Black, Ashley P and Frontera, Carlos and Torres, Arturo and Recio-Poo, Miguel and Rozier, Patrick and Forero-Saboya, Juan D and Fauth, Fran{\c{c}}ois and Urones-Garrote, Esteban and Arroyo-de Dompablo, M Elena and Palac{\'\i}n, M Rosa},
  journal={Energy Storage Materials},
  volume={47},
  pages={354--364},
  year={2022},
  publisher={Elsevier}
}

@article{tekliye2026geometry,
  title={Geometry-based Discovery of Calcium Battery Cathodes Accelerated by Foundational Machine-Learned Models},
  author={Tekliye, Dereje Bekele and Bheemaguli, Achinthya Krishna and Gautam, Gopalakrishnan Sai},
  journal={arXiv preprint arXiv:2605.29029},
  year={2026}
}

@article{swathilakshmi2023performance,
  title={Performance of the r2SCAN functional in transition metal oxides},
  author={Swathilakshmi, S and Devi, Reshma and Sai Gautam, Gopalakrishnan},
  journal={Journal of chemical theory and computation},
  volume={19},
  number={13},
  pages={4202--4215},
  year={2023},
  publisher={ACS Publications}
}

@article{ong2013python,
  title={Python Materials Genomics (pymatgen): A robust, open-source python library for materials analysis},
  author={Ong, Shyue Ping and Richards, William Davidson and Jain, Anubhav and Hautier, Geoffroy and Kocher, Michael and Cholia, Shreyas and Gunter, Dan and Chevrier, Vincent L and Persson, Kristin A and Ceder, Gerbrand},
  journal={Computational Materials Science},
  volume={68},
  pages={314--319},
  year={2013},
  publisher={Elsevier}
}

@article{ewald1921berechnung,
  title={Die Berechnung optischer und elektrostatischer Gitterpotentiale},
  author={Ewald, Paul P},
  journal={Annalen der physik},
  volume={369},
  number={3},
  pages={253--287},
  year={1921},
  publisher={Wiley Online Library}
}

@article{hinuma2007phase,
  title={Phase transitions in the LiNi0. 5Mn0. 5O2 system with temperature},
  author={Hinuma, Yoyo and Meng, Ying S and Kang, Kisuk and Ceder, Gerbrand},
  journal={Chemistry of Materials},
  volume={19},
  number={7},
  pages={1790--1800},
  year={2007},
  publisher={ACS Publications}
}

@article{hinuma2008temperature,
  title={Temperature-concentration phase diagram of P 2-Na x CoO 2 from first-principles calculations},
  author={Hinuma, Yoyo and Meng, Ying S and Ceder, Gerbrand},
  journal={Physical Review B—Condensed Matter and Materials Physics},
  volume={77},
  number={22},
  pages={224111},
  year={2008},
  publisher={APS}
}

@article{van2000phase,
  title={Phase transformations and volume changes in spinel LixMn2O4},
  author={Van der Ven, A and Marianetti, C and Morgan, D and Ceder, G},
  journal={Solid State Ionics},
  volume={135},
  number={1-4},
  pages={21--32},
  year={2000},
  publisher={Elsevier}
}

@article{van2010linking,
  title={Linking the electronic structure of solids to their thermodynamic and kinetic properties},
  author={Van der Ven, Anton and Thomas, John C and Xu, Qingchuan and Bhattacharya, Jishnu},
  journal={Mathematics and computers in simulation},
  volume={80},
  number={7},
  pages={1393--1410},
  year={2010},
  publisher={Elsevier}
}

@article{devi2022effect,
  title={Effect of exchange-correlation functionals on the estimation of migration barriers in battery materials},
  author={Devi, Reshma and Singh, Baltej and Canepa, Pieremanuele and Sai Gautam, Gopalakrishnan},
  journal={npj Computational Materials},
  volume={8},
  number={1},
  pages={160},
  year={2022},
  publisher={Nature Publishing Group UK London}
}

@article{tekliye2022exploration,
  title={Exploration of nasicon frameworks as calcium-ion battery electrodes},
  author={Tekliye, Dereje Bekele and Kumar, Ankit and Weihang, Xie and Mercy, Thelakkattu Devassy and Canepa, Pieremanuele and Sai Gautam, Gopalakrishnan},
  journal={Chemistry of Materials},
  volume={34},
  number={22},
  pages={10133--10143},
  year={2022},
  publisher={ACS Publications}
}

@article{tekliye2024fluoride,
  title={Fluoride frameworks as potential calcium battery cathodes},
  author={Tekliye, Dereje Bekele and Gautam, Gopalakrishnan Sai},
  journal={Journal of Materials Chemistry A},
  volume={12},
  number={30},
  pages={18993--19007},
  year={2024},
  publisher={Royal Society of Chemistry}
}

@article{gummow2018calcium,
  title={Calcium-ion batteries: current state-of-the-art and future perspectives},
  author={Gummow, Rosalind J and Vamvounis, George and Kannan, M Bobby and He, Yinghe},
  journal={Advanced Materials},
  volume={30},
  number={39},
  pages={1801702},
  year={2018},
  publisher={Wiley Online Library}
}

@article{rong2015materials,
  title={Materials design rules for multivalent ion mobility in intercalation structures},
  author={Rong, Ziqin and Malik, Rahul and Canepa, Pieremanuele and Sai Gautam, Gopalakrishnan and Liu, Miao and Jain, Anubhav and Persson, Kristin and Ceder, Gerbrand},
  journal={Chemistry of Materials},
  volume={27},
  number={17},
  pages={6016--6021},
  year={2015},
  publisher={ACS Publications}
}

@article{wang2018electrolyte,
  author = {Wang, Da and Gao, Xiangwen and Chen, Yuhui and Jin, Liyu and Kuss, Christian and Bruce, Peter G.},
  title = {Plating and stripping calcium in an organic electrolyte},
  journal = {Nature Materials},
  volume = {17},
  number = {1},
  pages = {16--20},
  year = {2018},
  publisher={Nature Publishing Group}
}

@article{li2019towards,
  title={Towards stable and efficient electrolytes for room-temperature rechargeable calcium batteries},
  author={Li, Zhenyou and Fuhr, Olaf and Fichtner, Maximilian and Zhao-Karger, Zhirong},
  journal={Energy \& Environmental Science},
  volume={12},
  number={12},
  pages={3496--3501},
  year={2019},
  publisher={Royal Society of Chemistry}
}

@article{pu2020current,
  title={Current-density-dependent electroplating in Ca electrolytes: From globules to dendrites},
  author={Pu, Shengda D and Gong, Chen and Gao, Xiangwen and Ning, Ziyang and Yang, Sixie and Marie, John-Joseph and Liu, Boyang and House, Robert A and Hartley, Gareth O and Luo, Jun and others},
  journal={ACS Energy Letters},
  volume={5},
  number={7},
  pages={2283--2290},
  year={2020},
  publisher={ACS Publications}
}

@article{shyamsunder2019reversible,
  title={Reversible calcium plating and stripping at room temperature using a borate salt},
  author={Shyamsunder, Abhinandan and Blanc, Lauren E and Assoud, Abdeljalil and Nazar, Linda F},
  journal={ACS Energy Letters},
  volume={4},
  number={9},
  pages={2271--2276},
  year={2019},
  publisher={ACS Publications}
}

@article{tekliye2024accuracy,
  title={Accuracy of metaGGA functionals in describing transition metal fluorides},
  author={Tekliye, Dereje Bekele and Sai Gautam, Gopalakrishnan},
  journal={Physical Review Materials},
  volume={8},
  number={9},
  pages={093801},
  year={2024},
  publisher={APS}
}

@article{jeon2020reversible,
  title={Reversible Calcium-Ion Insertion in NASICON-Type NaV2(PO4)3},
  author={Jeon, Boosik and Heo, Jongwook W. and Hyoung, Jooeun and Kwak, Hunho H. and Lee, Dongmin M. and Hong, Seung-Tae},
  year={2020},
  month={10},
  day={27},
  journal={Chemistry of Materials},
  volume={32},
  number={20},
  pages={8772--8780},
  publisher={American Chemical Society}
}

@article{kim2020high,
  title={High-Voltage Phosphate Cathodes for Rechargeable Ca-Ion Batteries},
  author={Kim, Sanghyeon and Yin, Liang and Lee, Myeong Hwan and Parajuli, Prakash and Blanc, Lauren and Fister, Timothy T. and Park, Haesun and Kwon, Bob Jin and Ingram, Brian J. and Zapol, Peter and Klie, Robert F. and Kang, Kisuk and Nazar, Linda F. and Lapidus, Saul H. and Vaughey, John T.},
  year={2020},
  month={10},
  day={9},
  journal={ACS Energy Letters},
  volume={5},
  number={10},
  pages={3203--3211},
  publisher={American Chemical Society}
}

@article{kresse1993ab,
  title={Ab initio molecular dynamics for liquid metals},
  author={Kresse, Georg and Hafner, J{\"u}rgen},
  journal={Physical review B},
  volume={47},
  number={1},
  pages={558},
  year={1993},
  publisher={APS}
}

@article{kresse1996efficient,
  title={Efficient iterative schemes for ab initio total-energy calculations using a plane-wave basis set},
  author={Kresse, Georg and Furthm{\"u}ller, J{\"u}rgen},
  journal={Physical Review B},
  volume={54},
  number={16},
  pages={11169},
  year={1996},
  publisher={APS}
}

@article{sun2015strongly,
  title={Strongly constrained and appropriately normed semilocal density functional},
  author={Sun, Jianwei and Ruzsinszky, Adrienn and Perdew, John P},
  journal={Physical Review Letters},
  volume={115},
  number={3},
  pages={036402},
  year={2015},
  publisher={APS}
}

@article{dudarev1998electron,
  title={Electron-energy-loss spectra and the structural stability of nickel oxide: An {LSDA+\textit{U}} study},
  author={Dudarev, Sergei L and Botton, Gianluigi A and Savrasov, Sergey Y and Humphreys, CJ and Sutton, Adrian P},
  journal={Physical Review B},
  volume={57},
  number={3},
  pages={1505},
  year={1998},
  publisher={APS}
}

@article{anisimov1991band,
  title={Band theory and Mott insulators: Hubbard \textit{U} instead of Stoner \textit{I}},
  author={Anisimov, Vladimir I and Zaanen, Jan and Andersen, Ole K},
  journal={Physical Review B},
  volume={44},
  number={3},
  pages={943},
  year={1991},
  publisher={APS}
}

@article{gautam2018evaluating,
  title={Evaluating transition metal oxides within {DFT-SCAN and SCAN+\textit{U}} frameworks for solar thermochemical applications},
  author={Gautam, Gopalakrishnan Sai and Carter, Emily A},
  journal={Physical Review Materials},
  volume={2},
  number={9},
  pages={095401},
  year={2018},
  publisher={APS}
}

@article{long2020evaluating,
  title={Evaluating optimal \textit{U} for 3\textit{d} transition-metal oxides within the {SCAN+\textit{U}} framework},
  author={Long, Olivia Y and Gautam, Gopalakrishnan Sai and Carter, Emily A},
  journal={Physical Review Materials},
  volume={4},
  number={4},
  pages={045401},
  year={2020},
  publisher={APS}
}

@article{kresse1999ultrasoft,
  title={From ultrasoft pseudopotentials to the projector augmented-wave method},
  author={Kresse, Georg and Joubert, Daniel},
  journal={Physical Review B},
  volume={59},
  number={3},
  pages={1758},
  year={1999},
  publisher={APS}
}

@article{blochl1994improved,
  title={Improved tetrahedron method for Brillouin-zone integrations},
  author={Bl{\"o}chl, Peter E and Jepsen, Ove and Andersen, Ole Krogh},
  journal={Physical Review B},
  volume={49},
  number={23},
  pages={16223},
  year={1994},
  publisher={APS}
}

@article{monkhorst1976special,
  title={Special points for Brillouin-zone integrations},
  author={Monkhorst, Hendrik J and Pack, James D},
  journal={Physical Review B},
  volume={13},
  number={12},
  pages={5188},
  year={1976},
  publisher={APS}
}

@article{hellenbrandt2004inorganic,
  title={The inorganic crystal structure database ({ICSD})—present and future},
  author={Hellenbrandt, Mariette},
  journal={Crystallography Reviews},
  volume={10},
  number={1},
  pages={17--22},
  year={2004},
  publisher={Taylor \& Francis}
}

@article{sanchez1984generalized,
  title={Generalized cluster description of multicomponent systems},
  author={Sanchez, Juan M and Ducastelle, Francois and Gratias, Denis},
  journal={Physica A: Statistical Mechanics and its Applications},
  volume={128},
  number={1-2},
  pages={334--350},
  year={1984},
  publisher={Elsevier}
}

@article{puchala2013thermodynamics,
  title={Thermodynamics of the Zr-O system from first-principles calculations},
  author={Puchala, B and Van der Ven, A},
  journal={Physical Review B—Condensed Matter and Materials Physics},
  volume={88},
  number={9},
  pages={094108},
  year={2013},
  publisher={APS}
}

@article{puchala2023casm,
  title={CASM—A software package for first-principles based study of multicomponent crystalline solids},
  author={Puchala, Brian and Thomas, John C and Natarajan, Anirudh Raju and Goiri, Jon Gabriel and Behara, Sesha Sai and Kaufman, Jonas L and Van der Ven, Anton},
  journal={Computational Materials Science},
  volume={217},
  pages={111897},
  year={2023},
  publisher={Elsevier}
}

@article{van2018first,
  title={First-principles statistical mechanics of multicomponent crystals},
  author={Van der Ven, Anton and Thomas, John C and Puchala, Brian and Natarajan, Anirudh Raju},
  journal={Annual Review of Materials Research},
  volume={48},
  number={1},
  pages={27--55},
  year={2018},
  publisher={Annual Reviews}
}

@article{henkelman2000improved,
  title={Improved tangent estimate in the nudged elastic band method for finding minimum energy paths and saddle points},
  author={Henkelman, Graeme and J{\'o}nsson, Hannes},
  journal={The Journal of chemical physics},
  volume={113},
  number={22},
  pages={9978--9985},
  year={2000},
  publisher={American Institute of Physics}
}

@article{sheppard2008optimization,
  title={Optimization methods for finding minimum energy paths},
  author={Sheppard, Daniel and Terrell, Rye and Henkelman, Graeme},
  journal={The Journal of chemical physics},
  volume={128},
  pages={},
  number={13},
  year={2008},
  publisher={AIP Publishing}
}

@article{perdew1996generalized,
  title={Generalized gradient approximation made simple},
  author={Perdew, John P and Burke, Kieron and Ernzerhof, Matthias},
  journal={Physical review letters},
  volume={77},
  number={18},
  pages={3865},
  year={1996},
  publisher={APS}
}

@article{niazi2009single,
  title={Single-crystal growth, crystallography, magnetic susceptibility, heat capacity, and thermal expansion of the antiferromagnetic S= 1 chain compound CaV 2 O 4},
  author={Niazi, A and Bud’ko, Sergey L and Schlagel, Deborah L and Yan, JQ and Lograsso, Thomas A and Kreyssig, A and Das, S and Nandi, S and Goldman, AI and Honecker, Andreas and others},
  journal={Physical Review B—Condensed Matter and Materials Physics},
  volume={79},
  number={10},
  pages={104432},
  year={2009},
  publisher={APS}
}

@article{metropolis1953equation,
  title={Equation of state calculations by fast computing machines},
  author={Metropolis, Nicholas and Rosenbluth, Arianna W and Rosenbluth, Marshall N and Teller, Augusta H and Teller, Edward},
  journal={The journal of chemical physics},
  volume={21},
  number={6},
  pages={1087--1092},
  year={1953},
  publisher={American Institute of Physics}
}

@article{deng2020phase,
  title={Phase behavior in rhombohedral NaSiCON electrolytes and electrodes},
  author={Deng, Zeyu and Sai Gautam, Gopalakrishnan and Kolli, Sanjeev Krishna and Chotard, Jean-No{\"e}l and Cheetham, Anthony K and Masquelier, Christian and Canepa, Pieremanuele},
  journal={Chemistry of Materials},
  volume={32},
  number={18},
  pages={7908--7920},
  year={2020},
  publisher={ACS Publications}
}

@article{wang2022phase,
  title={Phase stability and sodium-vacancy orderings in a NaSICON electrode},
  author={Wang, Ziliang and Park, Sunkyu and Deng, Zeyu and Carlier, Dany and Chotard, Jean-No{\"e}l and Croguennec, Laurence and Gautam, Gopalakrishnan Sai and Cheetham, Anthony K and Masquelier, Christian and Canepa, Pieremanuele},
  journal={Journal of Materials Chemistry A},
  volume={10},
  number={1},
  pages={209--217},
  year={2022},
  publisher={Royal Society of Chemistry}
}

@article{lee2024thermodynamics,
  title={Thermodynamics of Sodium--Lead Alloys for Negative Electrodes from First-Principles},
  author={Lee, Damien KJ and Deng, Zeyu and Sai Gautam, Gopalakrishnan and Canepa, Pieremanuele},
  journal={Chemistry of Materials},
  volume={36},
  number={14},
  pages={6831--6837},
  year={2024},
  publisher={ACS Publications}
}

@article{liu2014capturing,
  title={Capturing metastable structures during high-rate cycling of LiFePO4 nanoparticle electrodes},
  author={Liu, Hao and Strobridge, Fiona C and Borkiewicz, Olaf J and Wiaderek, Kamila M and Chapman, Karena W and Chupas, Peter J and Grey, Clare P},
  journal={Science},
  volume={344},
  number={6191},
  pages={1252817},
  year={2014},
  publisher={American Association for the Advancement of Science}
}

@article{xu2025understanding,
  title={Understanding and Optimizing Li Substitution in P2-Type Sodium Layered Oxides for Sodium-Ion Batteries},
  author={Xu, Mingfeng and Gammaitoni, Giovanni and H{\"a}fner, Michael and Villalobos-Portillo, Eduardo and Marini, Carlo and Bianchini, Matteo},
  journal={Advanced Functional Materials},
  pages={2425499},
  year={2025},
  publisher={Wiley Online Library}
}

@article{van2002self,
  title={Self-driven lattice-model Monte Carlo simulations of alloy thermodynamic properties and phase diagrams},
  author={Van De Walle, Axel and Asta, Mark},
  journal={Modelling and Simulation in Materials Science and Engineering},
  volume={10},
  number={5},
  pages={521--538},
  year={2002}
}

@article{batatia2025design,
  title={The design space of E (3)-equivariant atom-centred interatomic potentials},
  author={Batatia, Ilyes and Batzner, Simon and Kov{\'a}cs, D{\'a}vid P{\'e}ter and Musaelian, Albert and Simm, Gregor NC and Drautz, Ralf and Ortner, Christoph and Kozinsky, Boris and Cs{\'a}nyi, G{\'a}bor},
  journal={Nature Machine Intelligence},
  volume={7},
  number={1},
  pages={56--67},
  year={2025},
  publisher={Nature Publishing Group UK London}
}

@article{rhodes2025orb,
  title={Orb-v3: atomistic simulation at scale},
  author={Rhodes, Benjamin and Vandenhaute, Sander and {\v{S}}imkus, Vaidotas and Gin, James and Godwin, Jonathan and Duignan, Tim and Neumann, Mark},
  journal={arXiv preprint arXiv:2504.06231},
  year={2025}
}

@article{devi2026leveraging,
  title={Leveraging transfer learning for accurate estimation of ionic migration barriers in solids},
  author={Devi, Reshma and Butler, Keith T and Sai Gautam, Gopalakrishnan},
  journal={npj Computational Materials},
  year={2026},
  publisher={Nature Publishing Group UK London}
}

@article{malik2011kinetics,
  title={Kinetics of non-equilibrium lithium incorporation in LiFePO4},
  author={Malik, Rahul and Zhou, Fei and Ceder, Gerbrand},
  journal={Nature materials},
  volume={10},
  number={8},
  pages={587--590},
  year={2011},
  publisher={Nature Publishing Group UK London}
}

@article{zhou2006configurational,
  title={Configurational Electronic Entropy and the Phase Diagram of Mixed-Valence Oxides:<? format?> The Case of Li x FePO 4},
  author={Zhou, Fei and Maxisch, Thomas and Ceder, Gerbrand},
  journal={Physical review letters},
  volume={97},
  number={15},
  pages={155704},
  year={2006},
  publisher={APS}
}

\end{document}